\documentclass[referee]{raa}            % referee version: for submission

\usepackage{graphicx,times}             %for PS/EPS graphics inclusion, new
\usepackage{hyperref}
\usepackage{filecontents}
\usepackage{amssymb}
\usepackage{xcolor}
\usepackage{natbib}
\usepackage{amsmath}
\usepackage{soul}
\usepackage[a4paper, total={6in, 8in}]{geometry}

\hypersetup{colorlinks,linkcolor={blue},citecolor={blue},urlcolor={red}}

\begin{document}

   \title{Effect of magnetic flux advection on the dynamics of shock in accretion flow around a rotating black hole
%\footnotetext{$*$ Supported by the National Natural Science Foundation of China.}
}
%   \subtitle{I. Place Your Subtitle Here}

   \volnopage{Vol.0 (200x) No.0, 000--000}      %%preserved for Editor. DOn't remove!
   \setcounter{page}{1}          %%starting page, preserved for Editor. DOn't remove!

   \author{Biplob Sarkar
      \inst{1}
   \and Anjali Rao
      \inst{2}
   }

   \institute{Faculty of Science \& Technology, The ICFAI University Tripura, Agartala, 
799210, India; {\it biplobsarkar@iutripura.edu.in}\\
        \and
             Department of Physics and Astronomy, University of Southampton, Highfield, Southampton SO17 1BJ\\
   }

   \date{Received~~2016 month day; accepted~~20016~~month day}

\abstract{ We investigate the dynamical behaviour of a magnetized, dissipative,  accretion flow around a rapidly rotating black hole. We solve the magnetohydrodynamic equations and calculate the transonic accretion solutions which may contain discontinuous shock transitions. We investigate the effect of $\zeta-$ parameter (parametrizing the radial variation of the toroidal magnetic flux advection rate) on the dynamical behaviour of shocks. For a rapidly rotating black hole and for fixed injection parameters at the outer edge, we show that stationary shocks are sustained in the global magnetized accretion solutions for a wide range of $\zeta$ and accretion rate ($\dot{m}$). To investigate the observational implications, we consider dissipative shocks and estimate the maximum accessible energy from the post-shock corona (PSC) for nine stellar mass black hole candidates. We compare this with the observed radio jet kinetic power reported in the literature, whenever available. We find close agreement between the estimated values from our model with those reported in the literature. 
}

   \authorrunning{Sarkar \& Rao}            %author_head in even pages
   \titlerunning{Magnetized accretion flow around a black hole}  % title_head in odd pages

   \maketitle
\section{Introduction}
Magnetic fields have a significant effect in a variety of astrophysical processes. Specifically in an accretion disc, there must be transport of angular momentum outwards so that matter can fall in at the central object. In this process, the utility of magnetic fields in the accretion disc was first discussed by \citet{Shakura1973}. By considering the magnetic field to be turbulent, the authors integrated its role to the viscosity in the accretion disc which is well-known as the $\alpha-$parameter these days. Moreover, magnetic fields play a significant role in the origin of bi-directional jet from the inner region of the accretion disc around a black hole (BH). \citet{Blandford1977} proposed that relativistic jet may be powered by the rotational energy of the BH via torque exerted by magnetic field lines threading the BH horizon. In an alternative mechanism, \citet{Blandford1982} suggested that a spinning BH actually causes the ordered magnetic field in the accretion disc to corotate with the accreting matter and this drives quasi-relativistic jets via the release of gravitational potential energy. Again, it was shown by \citet{Balbus1991} that the Magneto-Rotational Instability (MRI) may give rise to the disc viscosity, where even a weak magnetic field can make the accretion flow go turbulent leading to the outward transport of angular momentum. Recent 3D magnetohydrodynamic (MHD) simulations assuming azimuthal magnetic fields have further demonstrated  the formation of a magnetically supported accretion disc around a BH \citep{machida06,johansen2008}. Optically thin accretion disc supported by magnetic pressure can account for the ``Bright/Hard'' state in BH candidates which is observed when the accretion disc undergoes transition from a Low/Hard state to the High/Soft state \citep{oda10,oda12}.

In the model of magnetized accretion flow around BHs considered by \citet{oda07,oda12, Sarkar-Das2016, Das-Sarkar2018}, the parameter $\zeta$ parametrizes the radial variation of the toroidal magnetic flux advection rate ($\dot{\Phi}$). \citet{oda12} showed that the change in $\zeta$ results in modification of $\dot{\Phi}$ which is responsible for the state transitions in BH candidates (BHCs). When $\dot{\Phi}$ is relatively high and $\dot{M}$ surpasses a critical limit for the trigger of cooling instability, the advection dominated accretion flow (ADAF) \citep{Ichimaru1977,Narayan1994,Narayan1995,Abramowicz1995} develops towards a magnetic pressure dominated disc. In this situation, \citet{oda12} proposed that the transition of the accretion disc would be bright hard-to-soft. While for a low $\dot{\Phi}$, when $\dot{M}$ surpasses the limit, the ADAF advances towards an optically thick disc. In this situation, \cite{oda12} suggested that the transition would be dark hard-to-soft, which happens at under 0.1 $L_{\rm Edd}$ \citep{oda10}.

Another significant aspect in the study of accretion flow around a BH is the existence of global accretion 
solutions containing a shock, which have a special importance. Such shocked accretion solutions occur when the supersonic accretion flow approaching towards a BH experiences the centrifugal barrier in the vicinity of the BH. Due to the centrifugal repulsion, the supersonic flow may undergo a shock transition to become subsonic and the post-shock corona (hereafter PSC) \citep{Aktar-etal15,Sarkar-Das2016,Dihingia2019} becomes hot, dense and puffed up \citep{Fukue87,Lu-Yuan98,dcnc01,Gu-Lu04,Das_Choi2009,Chattopadhyay-Chakrabarti2011,Sarkar-Das2018,Dihingia-etal2018}. Shocked accretion solutions around BHs have the potential to explain many intriguing features of BH candidates like the origin of outflows,  jets \citep[and references therein]{Das-etal14a,Das-etal14b} and the quasi-periodic oscillations (QPO's) in the hard X-ray spectrum of BHCs \citep{Molteni-etal96,Chakrabarti-etal2004}. 

In recent years, \citet{sarkar_2015, Sarkar-Das2016, Sarkar-et-al2018, Sarkar-Das2018} have extensively investigated the implication of shocks in accretion flow around non-rotating BHs, where the disc is threaded by toroidal magnetic fields. \citet{Das-Sarkar2018} extended the study to investigate the combined effect of viscosity and toroidal magnetic fields on the dynamical structure of the global accretion flow around rotating BHs. In these works, the authors assumed $\zeta$ = 1 throughout, as a representative case. However, in general it is expected that the variation of $\zeta$ will have significant effect on the radial structure of the disc. In particular, \citet{oda07} showed that the variation of $\zeta$ affects the temperature, plasma$-\beta$, accretion rate and optical depth in the disc. Since $\zeta$ will affect the angular momentum transport in the disc as well, the dynamics of shock is also likely to be governed by the value of $\zeta$. In this paper, we aim to investigate the effect of $\zeta$ on the dynamical behaviour of shocks in a viscous axisymmetric magnetized accretion flow around a rapidly spinning BH. To mimic the space-time geometry around a spinning BH, we use the pseudo Kerr potential intoduced by \citet{Chak-Mondal2006}. The adopted pseudo potential in this work represents a rotating black hole with spin parameter $a_s \lesssim 0.8$. The relevant works related to other form of pseudo potentials commonly used by the researchers are \citet{Chakrabarti1992,Lovas1998,Artemova1996,Semerak1999,Mukhopadhyay2002,Ivanov2005,Ghosh2007,Ghosh2014,Karas2014}. Very recently \citet{Dihingia2018}, suggested another effective pseudo-potential which exactly describes the space-time geometry around a Kerr black hole for the range of spin $0\leq a_s <1$. The focus of the present work is to determine the critical value of $\zeta$ which would sustain standing shocks in the magnetized accretion solutions around a rapidly rotating BH. Also, we show that a wide range of $\zeta$ sustains standing shocks in the accretion flow. Further for fixed flow parameters at the outer edge, we show that for a flow with higher $\zeta$, the critical accretion rate ($\dot{m}^{\rm cri}$) that sustains shock is lower. Finally, we make use of the present formalism to calculate the maximum shock luminosity ($\mathcal{L}_{\rm shock}^{\rm max}$) that corresponds to the maximum energy dissipated across the shock. This is done for nine stellar mass BH sources, where the mass, accretion rate and spin of the individual sources are taken from the literature. We compare $\mathcal{L}_{\rm shock}^{\rm max}$ with the reported values of jet kinetic power of the sources in the literature, whenever available, and we find close agreement between the two.

In the following, we first present the model assumptions and governing equations and the methodology to solve 
them (Section 2). In Section 3, we present the shocked global accretion solutions, the properties of shock and 
the critical values of $\zeta^{\rm cri}$ and $\dot{m}^{\rm cri}$ for shock. 
Next, in Section 4, we compute the maximum shock luminosity from our model and compare it for several stellar mass BH
sources. Finally, we present the summary of the work in Section 5. 

\section{Model for the accretion flow}

The structure of magnetic fields in the accretion disc is considered to be the same as described 
in \citet{oda12}. On the basis of the results obtained from 3D global as well as local MHD 
simulations of accretion flow around BHs, we consider the magnetic fields 
in the accretion disc to be turbulent and dominated by the toroidal component \citep{hirose2006,
machida06,johansen2008}. According to the results of these simulations, the magnetic fields in 
the disc are decomposed into the mean fields, denoted by ${\bf{B}} = (0,<B_{\phi}>,0)$, and 
the fluctuating fields, $\delta {\bf{B}}= (\delta {B}_{r}, \delta {B}_{\phi},\delta {B}_{z})$. 
Here, `$<>$' expresses the azimuthal average of any quantity. The fluctuating components of 
magnetic fields disappear on azimuthally averaging ($<\delta {\bf{B}}> = 0$). Moreover, the 
azimuthal component of the magnetic field is much more significant as compared to the 
radial and vertical components, $\mid <B_{\phi}> + \delta B_{\phi}\mid \gg \mid \delta B_{r} \mid ~{\rm and} 
\mid \delta B_{z}\mid$ (see Fig. 2 of \citet{oda12}). Eventually, the azimuthally averaged form of the magnetic 
field is obtained as $<{\bf{B}}>=<B_{\phi}>\hat{\phi}$ \citep{oda07}.

\subsection{Governing Equations}

In this work, we consider a steady, thin and axisymmetric magnetized accretion flow around a rotating BH. 
Throughout the paper, we express the radial coordinate ($x$) in units of $r_g = GM_{\rm BH}/c^2$, where 
$G$ is the universal gravitational constant, $M_{\rm BH}$ is the mass of the BH and $c$ is the speed of light. 
Also, flow velocity is measured in units of $c$ and time is measured in units of $GM_{\rm BH}/c^3$. 
We assume the ($x$, $\phi$, $z$) coordinate system where the accretion flow is confined in the $x-\phi$ plane 
and the BH is located at the center of the coordinate system. 

We denote the dynamical flow variables, namely, radial velocity, sound speed, mass density, specific angular momentum, temperature, adiabatic index,  specific entropy, gas pressure and magnetic pressure of the flow by the symbols $u$, $c_s$, $\rho$, $\lambda$, $T$, $\gamma$, $s$, $p_{\rm gas}$ and $p_{\rm mag}$ respectively. Hence the mass flux conservation equation, radial momentum equation, azimuthal momentum equation and the entropy generation equation are, respectively,
$$
\dot{M}=2\pi u\Sigma x,
\eqno(1)
$$
$$
{u\frac{du}{dx} + \frac{1}{\rho}\frac{dp_{\rm tot}}{dx} 
+ \frac{d\Psi_{\rm eff}}{dx} + \frac{\left<B_{\phi} ^2\right>}{4\pi x \rho} = 0},
\eqno(2)
$$
$$
{u \Sigma x}\frac{d\lambda}{dx}+ \frac{d}{dx}(x^2\mathbb{T}_{x\phi}) = 0,
\eqno(3)
$$
$$
u T\Sigma \frac {ds}{dx}=\frac{Hu}{1 - \gamma}
\left(\frac{\gamma p_{\rm gas}}{\rho}\frac{d\rho}{dx} - \frac{dp_{\rm gas}}{dx}\right)=\mathcal{Q}^- - \mathcal{Q}^+.
\eqno(4)
$$
Here, $\dot{M}$ represents the mass accretion rate and $\Sigma$ specifies the mass density of the 
flow averaged in the vertical direction \citep{Matsumoto-etal1984}. It might be noticed that in 
the present work, we consistently regard the direction of inward radial velocity as positive. Also, 
$p_{\rm tot}$ is the total pressure of the flow which is considered to be 
$p_{\rm tot} = p_{\rm gas} + p_{\rm mag}$. The gas pressure inside the disc is obtained as, 
$p_{\rm gas} = R\rho T/\mu$ where $R$ is the gas constant and $\mu$ is the mean molecular 
weight. We consider $\mu = 0.5$ for a fully ionized hydrogen gas. Also, the magnetic pressure 
of the flow is obtained as, $p_{\rm mag} = <B_{\phi}^2>/8\pi$. We define the magnetic parameter 
of the flow as $\beta = p_{\rm gas}/p_{\rm mag}$. Using the definition of magnetic 
parameter, we obtain the total pressure of the flow as $p_{\rm tot} = p_{\rm gas} (1 + 1/\beta)$. 
Moreover, $\mathbb{T}_{x\phi}$ denotes the $x\phi$ component of the Maxwell stress and this component 
dominates the vertically integrated total stress in the flow. When velocity in the radial direction 
is significant in the accretion flow, $\mathbb{T}_{x\phi}$ is obtained as,
$$
\mathbb{T}_{x\phi} = \frac{<B_{x}B_{\phi}>}{4\pi}H = -\alpha_{T}(W + \Sigma u^2),
\eqno(5)
$$
where $H$, $\alpha_T$ and $W$, respectively, denote the disc half-thickness, constant of proportionality 
and the vertically integrated pressure of the flow \citep{Matsumoto-etal1984}. Along the same line as the seminal work of \citet{Shakura1973}, 
we consider $\alpha_T$ to remain constant everywhere along the flow. In the event that 
$u$ is insignificant, such as for a Keplerian flow, equation (5) restores to the 
`$\alpha$-model' \citep{Shakura1973}.

In left-hand side of equation (2), $\Psi_{\rm eff}$ is the effective potential around a spinning BH suggested by 
\citet{Chak-Mondal2006}. The form of $\Psi_{\rm eff}$ is given as,
$$
\Psi_{\rm eff} = - \frac{{\mathcal J} + \sqrt{{\mathcal J^2} - 4{\mathcal I}{\mathcal K}}}{2{\mathcal I}},
$$
where
$$
{\mathcal I} = \frac{\varepsilon^2 \lambda^2}{2x^2},
$$
$$
{\mathcal J} = -1 + \frac{\varepsilon^2 \varphi \lambda r^2}{x^2} + \frac{2\lambda a_s}{x r^2},
$$
$$
{\mathcal K} = 1 - \frac{1}{r - r_0} + \frac{2 a_s\varphi}{x} + \frac{\varepsilon^2 \varphi^2 r^4}{2x^2}.
$$
Here, radial distance in the cylindrical coordinates is represented by $x$ and in spherical 
coordinates it is denoted by $r$. In addition, $\lambda$ is the specific angular momentum 
of the flow. Further, $r_0 = 0.085a_s^2 + 0.97a_s + 0.04$, $\varphi = 2a_s /(2a_s^2 +a_s^2 x+x^3 )$ 
and $\varepsilon^2 = (a_s^2 - 2x + x^2 )/(2a_s^2 /x +a_s^2 + x^2)$, where $\varepsilon$ is 
the redshift factor and $a_s$ represents the spin of the BH. The pseudo 
potential used in this work adequately describes the space time geometry around a spinning 
BH for $a_s \lesssim 0.8$ \citep{Chak-Mondal2006}.

Assuming vertical hydrostatic equilibrium, we calculate the half-thickness of the disc ($H$)
as, $H(x)$ = $c_s \sqrt{{x}/({\gamma {{\Psi_{r}^{'}}}})}$ where 
$\Psi_{r}^{'}$ = $\left(\frac{\partial \Psi_{\rm eff}}{\partial r}\right)$ $_{z << x}$, 
$z$ indicates vertical scale height in the cylindrical coordinate system and $r =\sqrt {z^2 + x^2}$ 
\citep{Das-etal10,Aktar-etal15}. Here, the adiabatic sound speed is determined as $c_s=\sqrt {\gamma p_{\rm tot}/\rho}$. In this work, we accept $\gamma$ to remain 
constant throughout the flow and fix $\gamma=4/3$ (ultra-relativistic flow) for the analysis that follows.

Equation (4) is the entropy generation equation, where the cooling rate and heating rate of the flow are represented by $\mathcal{Q}^-$ and $\mathcal{Q}^+$, respectively. 3D MHD simulations have revealed that the dominant process contributing to the heating of the disc is the dissipation of magnetic energy via the magnetic reconnection mechanism \citep{machida06,hirose2006,Krolik2007}. In view of this, the heating rate is given by,
$$
\mathcal{Q}^{+} = \frac{<B_{x}B_{\phi}>}{4\pi} x H \frac{d\Omega}{dx} = 
-\alpha_{T}(W + \Sigma u^2) x \frac{d\Omega}{dx},
\eqno(6)
$$
where $\Omega$ represents the flow's angular velocity.

Due to high temperature at the inner region of the accretion disc, it is imperative that the plasma in this region be treated as two-temperature. In such a case, the electrons in the flow can lose energy via synchrotron and bremsstrahlung emission as well as through the inverse Comptonization effects. However, in the present work, we have approximated the plasma to be single temperature. Thus, the inverse Comptonization of the electrons has been disregarded since inclusion of this process requires a two-temperature analysis. Also, due to smaller size of the accretion disc in case of stellar mass BH candidates, the accretion disc is strongly threaded by magnetic fields. Hence, synchrotron emission by the electrons is expected to be dominant over the bremsstrahlung process \citep{Chatt-Chak2002,Rajesh2010}. Due to the above considerations, we consider only synchrotron cooling to be active in the flow. Accordingly, the synchrotron cooling rate is given as \citep{Shapiro-Teukolsky83},
$$
\mathcal{Q}^-= \frac{\widetilde{S}{c_s}^5\rho H}{u} \sqrt{\frac{\Psi_{r}^{'}}{x^3}} {\beta^2}{(\beta + 1)^{-3}},
\eqno(7)
$$
with,
$$
\widetilde{S}= 1.4827 \times 10^{18} \frac{ {\dot m} \mu^2 e^4}{I_n m_e^3\gamma^{5/2}}
G^{-1}M_{\odot}^{-1}c^{-3},
$$ 
where $e$ denotes the charge of the electron and $m_e$ represents the mass of the electron. Also, $\dot{m}$ 
signifies the accretion rate in Eddington units 
(${\dot M}_{\rm Edd} = 1.39 \times10^{17} \times M_{\rm BH}/M_{\odot}~{\rm g~s}^{-1}$). Here, we follow \citet{Rezzolla-2013,Straub-et-al-2014,Yuan2014} and use the efficiency factor of 1 to calculate the Eddington accretion rate. Moreover, $I_n = (2^n n!)^2/(2n + 1)!$ and $n (= 1/(\gamma - 1))$ indicates the polytropic index of the flow. We ignore any coupling between electron and ion and evaluate the temperature of electron in the flow using the relation $T_e = (\sqrt{m_e/m_i})T_i$ \citep{Chatt-Chak2002}. Here, $m_i$ indicates the mass of ion and $T_i$ refers to the ion temperature. It is to be noted that although we assume a single temperature flow, the temperature profile of ions and electrons is expected to become different at least at the inner disc region. It was indicated by \citet{Mahadevan1997} that the exchange of energy between ions and electrons via Coulomb collision is extremely ineffective in the accretion flow. Thus, the ions cannot attain thermal equilibrium with the electrons. The ions in principle retain the heat acquired via viscous dissipation in the disc. Again, since the relativistic electrons are much lighter than the non-relativistic ions, the electrons are able to radiate more effectively than the ions. Thus the accretion flow develops a two-temperature structure which is mostly prominent at the inner region of the disc.

Since we are dealing with hydromagnetic flows, as a consequence of induction equation in steady-state (for the current analysis), we obtain,
$$
{\bf \nabla} \times
\left({\vec{u}} \times <B_{\phi}>\hat{\phi} -{\frac{4\pi}{c}}\eta {\vec{j}}\right) = 0,
\eqno(8)
$$
where, $\vec {u}$ is the velocity vector,  $\vec {j}$ is the current density and $\eta$ is the resistivity of the flow. Equation (8) is now azimuthally averaged. On account of significantly large length scales in accretion disc, the Reynolds number achieves a very high value. As a result, we ignore the magnetic diffusivity term. Further, we also disregard the dynamo term for the present purpose. The resulting equation is then vertically integrated considering the vanishing of the azimuthally averaged toroidal magnetic fields at the surface of the disc. In view of these assumptions, the advection rate of toroidal magnetic flux is determined as,
$$
\dot{\Phi} = - \sqrt{4\pi}u {B}_{0} (x) H,
\eqno(9)
$$
where,
\begin{eqnarray*}
{B}_{0} (x) && = \langle {B}_{\phi} \rangle \left(x; z = 0\right)  \nonumber \\
&& = \frac{2^{5/4}{\pi}^{1/4}(R T)^{1/2}{\Sigma}^{1/2}}{\mu^{1/2} H^{1/2}{\beta}^{1/2}}
\end{eqnarray*}
represents the mean azimuthal toroidal magnetic field in the disc equatorial plane following \citet{oda07}. In the accretion disc, it is generally expected that $\dot{\Phi}$ will vary in the radial direction due to the diffusion of magnetic fields and the dynamo action. However, to avoid the complexity of determining the magnetic diffusivity term and dynamo term self consistently from the local variables, we follow the prescription of \citet{oda07} and adopt the parametric relation given by
$$
\dot{\Phi}\left(x; \zeta, \dot{M}\right) \equiv \dot{\Phi}_{\rm inj}
\left(\frac{x}{x_{\rm inj}} \right)^{-\zeta},
\eqno(10)
$$
where $\zeta$ represents a parameter describing the advection rate of magnetic flux. Further, $\dot{\Phi}_{\rm inj}$ denotes the advection rate of the toroidal magnetic field at the injection radius ($x_{\rm inj}$), equivalently the outer edge of the accretion disc. When $\zeta$$ = 0$, 
magnetic flux remains constant radially while, for $\zeta$$ > 0$, the magnetic flux increases with the decrease of radius.

\subsection{Transonic conditions and solution procedure}

Due to the essential transonic nature of BH accretion flows, the infalling matter is subsonic at the outer edge of the disc ($x_{\rm inj}$) and enters the BH supersonically \citep{Chak1990}. The location in the radial direction, where the flow behaviour changes smoothly from subsonic state to supersonic state is usually termed as a critical point. In the following, we solve the equations (1), (2), (3), (4), (9) and (10) simultaneously in order to perform the critical point analysis \citep[and references therein]{Sarkar-Das2016} and is given by,

$$
\frac {du}{dx}=\frac{\mathcal{N}}{\mathcal{D}},
\eqno(11)
$$
where the expressions for the numerator ($\mathcal{N}$) and denominator ($\mathcal{D}$) are provided in \citet{Das-Sarkar2018}. The expressions for the radial gradients of $c_s$, $\lambda$ and $\beta$ can also be found in \citet{Das-Sarkar2018}.

Now, the infalling matter must smoothly accrete onto the BH. This demands that the flow variables must have finite values everywhere along the flow. Thus, if the denominator $\mathcal{D}$ in equation (11) vanishes at any radial location, the numerator $\mathcal{N}$ must also simultaneously vanish there (${du}/{dx}= {0}/{0}$), in order to maintain a smooth solution. Such a radial location is known as a critical point ($x_c$). This gives us two critical point conditions as $\mathcal{N} = 0$ and $\mathcal{D} = 0$. The expression of Mach number ($M = u/c_s$) at $x_c$ is obtained by using the condition $\mathcal{D} = 0$, which is given as,
$$
M_c =\sqrt {\frac{-\varkappa_2 - \sqrt{\varkappa^2_2-4\varkappa_1 \varkappa_3}}{2\varkappa_1}},
\eqno(12)
$$
where
$$ 
\varkappa_1=2\alpha^2_{T} \gamma^2 I_n(1-\gamma)(1-2g)(\beta_c + 1) - \gamma^2[(1+\gamma)\beta_c + 3],
$$
$$
\varkappa_2=2\gamma(\gamma\beta_c + 2) + 4\alpha^2_{T} \gamma I_n g (1-g)(1-\gamma)(\beta_c + 1),
$$ 
$$
\varkappa_3=2\alpha^2_{T} g^2 I_n (1-\gamma)(\beta_c + 1),
$$
where, the parameter $g$ is defined as $g = I_{n + 1}/I_n$. To obtain the sound speed ($c_{\rm sc}$) at $x_c$, we use the condition $\mathcal{N}=0$ to get a cubic equation of the form,

$$
{\mathcal A_1}c_{\rm sc}^3 + {\mathcal A_2}c_{\rm sc}^2 + {\mathcal A_3}c_{\rm sc} +{\mathcal A_4}= 0 ,
\eqno(13)
$$
where
$$
{\mathcal A_1}=\widetilde{S}\sqrt{\frac{\Psi_{r}^{'}}{x_c^3}}\frac{\beta_c^{2}}{(\beta_c + 1)^{3}},
$$

$$
{\mathcal A_2} =  \frac {2\alpha^2_T I_n (\gamma M_c^2 + g)^2}
{x_c \gamma^2}-\frac{M_c^2(4+2\gamma\beta_c)}{2\gamma(1-\gamma)(\beta_c + 1)}\left(\frac{d{\rm ln}\Psi_{r}^{'}}{dx}\right) 
$$
$$
-\frac {2\alpha^2_T g I_n (\gamma M_c^2+g)}
{\gamma^2}\left(\frac{d{\rm ln}\Psi_{r}^{'}}{dx}\right)
$$
$$
 - \frac{2\{\beta_c(1+\gamma) + 3\}M_c^2}{\gamma(1-\gamma)(\beta_c + 1)^2 x_c} 
$$
$$
 + \frac{3(3+2\gamma\beta_c)M_c^2}{2\gamma(1-\gamma)(\beta_c + 1)x_c}+\frac {6\alpha^2_T g I_n(\gamma M_c^2 + g)}{ x_c \gamma^2}
$$
$$
-\frac {8\alpha^2_T g I_n(\gamma M_c^2 + g)}{\gamma^2 (\beta_c + 1)x_c}- \frac{(1-4\zeta)M_c^2}{2\gamma(\beta_c + 1)(1-\gamma) x_c},
$$

$$
{\mathcal A_3} = -\frac {4{\lambda_c} \alpha_T I_n M_c (\gamma M_c^2 + g)}{ x_c^2 \gamma},
$$

$$
{\mathcal A_4} = \left[\frac {\{\beta_c(1+\gamma)+3\}M_c^2}{(\gamma-1)(\beta_c + 1)}-\frac{4\alpha^2_T I_n g (\gamma M_c^2+g)}{\gamma} \right]\times\left(\frac{d\Psi_{r}^{'}}{dx}\right).
$$
%$$
%\times\left(\frac{d\Psi_{r}^{'}}{dx}\right).
%$$
Here, we represent the values of the flow variables estimated at $x_c$ with the subscript `c'.

Presently, we obtain the value of sound speed ($c_{\rm sc}$) at $x_c$ by supplying the accretion flow parameters in equation (13). Then the value of $u_c$ can be found using equation (12). The obtained values of $u_c$ and $c_{\rm sc}$ are supplied in Eq. (11) to analyze the characteristics of the critical points. At $x_c$, $du/dx$ possesses two particular values; one of which holds for the accretion solution and the other is for wind. If both the values of the velocity gradient at $x_c$ are real and of reverse sign, $x_c$ is called a `saddle' type critical point \citep{Chak1990,cd04}. Such a critical point is of unique significance since a global accretion solution goes through it only. In this work, since our objective is to explore the magnetized accretion flow around a BH, we consider solely the accretion solutions in the subsequent investigation.

\section{Results and discussion}

In order to acquire a global accretion solution, we need to perform the simultaneous solution of the coupled differential equations corresponding to the radial gradient of flow velocity ($u$), sound speed ($c_s$), angular momentum ($\lambda$) and plasma $\beta$. Additionally, the boundary values of $u$, $c_s$, $\lambda$, $\beta$ and $\dot{m}$ must be known at a given radial position ($x$). The Kerr parameter ($a_s$) and viscosity parameter ($\alpha_T$) also need to be specified. It is to be noted that, throughout the paper we express angular momentum ($\lambda$) in units of the Keplerian angular momentum $\lambda_K$(= $\sqrt{x^3/(x - 2)^2}$). Owing to the essential transonic nature of BH accretion solutions, flow must inevitably go through a critical point. Thus, it is suitable to supply the boundary flow variables at the critical point.  With this, we carry out integration of the equations corresponding to the radial gradient of $u$, $c_s$, $\lambda$ and $\beta$ from the critical point once towards the inside up to exactly outside the BH horizon and then outward up to the outer boundary of the disc. These two sections can be joined to yield a full global transonic accretion solution. As determined by the boundary parameters, the accretion flow has to cross one critical point at the minimum and feasibly more \citep[and references therein]{sarkar_2013}. A critical point which is located near the BH horizon is called as inner critical point ($x_{\rm in}$) and that which is located far from the BH horizon is known as outer critical point ($x_{\rm out}$). 

%%%%%%%%%%%%%%%%%%%%%%%%%%%%%
%          Fig 1          %
%%%%%%%%%%%%%%%%%%%%%%%%%%%%%
\begin{figure}[ht]
\centering
\includegraphics[width=10cm, height=10cm]{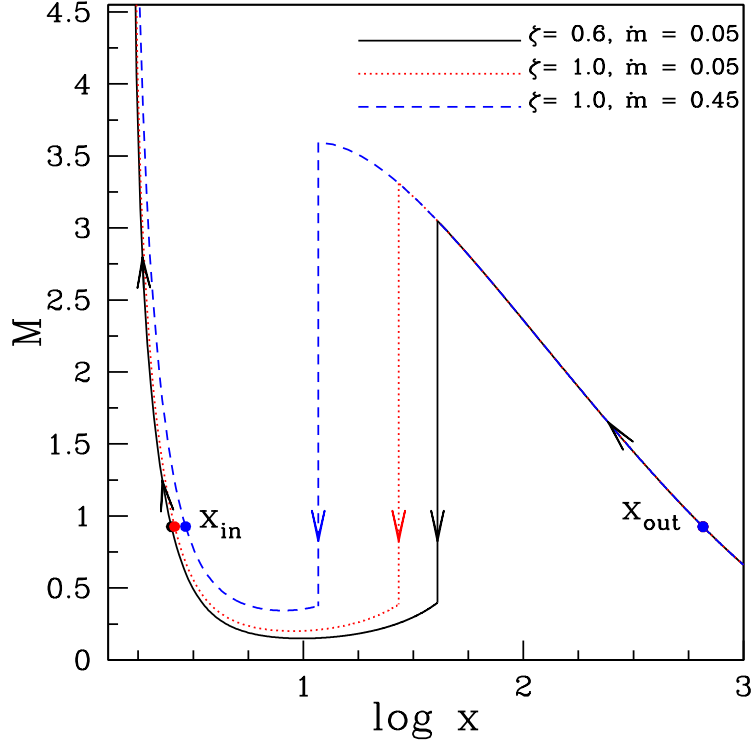}

\caption{Variation of Mach number as a function of logarithmic radial distance. The flows are injected from the outer edge $x_{\rm inj} = 1000$ with specific energy ${{E}}_{\rm inj}$$=8.9\times10^{-4}$, angular momentum $\lambda_{\rm inj}$ = 0.11014$\lambda_K$, $\beta_{\rm inj} = 10^5$, and viscosity $\alpha_T = 0.02$. The Kerr parameter is chosen as $a_s = 0.8$. Solid, dotted and dashed curves depict the results obtained for ($\zeta$, $\dot{m}$) = (0.6, 0.05), (1.0, 0.05) and (1.0, 0.45), respectively. The vertical arrows indicate the corresponding shock transitions positioned at $x_s$ = 40.72 (solid), $x_s$ = 27.12 (dotted) and $x_s$ = 11.68 (dashed). See text for details.}\label{fig:1}
\end{figure}
%%%%%%%%%%%%%%%%%%%%%%%%%%%%%%%%%%%%%%
To begin with, we explore the effect of the $\zeta$$-$parameter prescribing the magnetic flux advection rate on the dynamical structure of the accretion flow that contains shock wave. For this purpose, we fix the global parameters as $\alpha_T = 0.02$, $a_s = 0.8$ and inject matter subsonically from the outer edge of the disc at $x_{\rm inj} = 1000$ with local flow variables as $\lambda_{\rm inj}$ = 0.11014$\lambda_K$, $\beta_{\rm inj} = 10^5$ and ${{E}}_{\rm inj}$$=8.9\times10^{-4}$, respectively. The obtained results are depicted in Fig. 1, where the variation of Mach number ($M$) is plotted against the logarithmic radial distance. First we inject matter with accretion rate $\dot{m} = 0.05$ and $\zeta$$ = 0.6$. For these injection parameters, subsonic flow at the outer edge approaches towards the BH and becomes supersonic after crossing the outer critical point located at $x_{\rm out} = 654.52$. The supersonic flow proceeds towards the BH and sustains a virtual centrifugally repulsive barrier in the neighbourhood of the BH. Due to this obstruction, the accreting matter begins to pile up in the vicinity of the BH leading to the increase of flow density profile. However, this accumulation of matter cannot persist indefinitely and when the density scale length achieves a critical limit, the flow may suffer a discontinuous shock transition \citep{Frank-etal1992}. As a result of this, the ordered kinetic energy of the flow gets converted to disordered thermal energy. Whenever dynamically feasible, such a discontinuous shock jump is preferred by the laws of thermodynamics since the entropy of the flow increases after the shock \citep{Fukue87,Becker-Kazanas2001,Fukumura-Tsuruta2004}. The shock transition in a magnetized accretion flow is governed by the conservation of:\\
%\newline
\noindent (i) energy flux: ${{E}}_+$ = ${{E}}_-$\\
\noindent (ii) mass flux: $\dot{M}_+ = \dot{M}_-$\\
\noindent (iii) pressure balance: $W_+ + \Sigma_+u_+^2 = W_- + \Sigma_-u_-^2$\\
\noindent (iv) magnetic flux advection rate: $\dot{\Phi}_+$ = $\dot{\Phi}_-$
\newline
where `$+$' and `$-$' represent the quantities evaluated immediately after and before the shock 
transition ($x_s$) respectively \citep{Landau-Lifshitz1959,Sarkar-Das2016}. In the psuedo-Kerr 
geometry, we follow \citet{Das-Sarkar2018} to calculate the local specific energy of the flow as ${{E}}(x)$ 
$= u^2/2+c_s^2/(\gamma-1)+$ $\Psi_{\rm eff}$ $+<B_\phi ^2>/(4\pi \rho)$. For the 
present case, the shock is positioned at $x_s = 40.72$ as indicated with the 
solid vertical arrow. Soon after the shock transition,
flow gradually gains its radial velocity due to the gravitational attraction
and enters in to the BH supersonically after crossing the inner critical
point at $x_{\rm in} = 2.52$. When the magnetic flux advection
rate parameter is increased further as $\zeta$$=1.0$ keeping all the remaining flow parameters
unaltered, we find that shock front moves inward and settles down at
$x_s = 27.12$. In the figure, this is indicated using the dotted vertical arrow.
In reality, as $\zeta$ is increased, the magnetic fields carried with the accreting
matter are also enhanced that raise the synchrotron cooling efficiency. Since
the cooling is mostly effective at PSC, therefore post-shock pressure reduces
appreciably that eventually causes the shock front to move further towards the
BH in order to preserve the pressure balance across the shock. Next, we choose $\zeta$$=1.0$ 
and $\dot{m} = 0.45$ keeping rest of the flow parameters same.  Due to the increase of $\dot{m}$, 
the synchrotron cooling efficiency in the flow is additionally augmented and the 
shock front advances further towards the horizon. The shock now forms at $x_s = 11.68$ as shown with
the dashed vertical arrow. In all the cases, arrows indicate the overall direction
of flow motion towards the BH. Based on the above analysis, we
consequently point out that the $\zeta$$-$parameter prescribing the magnetic flux advection rate 
seems to play an important role in deciding the accretion disc dynamics including shock
waves, in addition to accretion rate ($\dot{m}$).
\newline
%%%%%%%%%%%%%%%%%%%%%%%%%%%%%
%          Fig 2          %
%%%%%%%%%%%%%%%%%%%%%%%%%%%%%
\begin{figure}[ht]
\centering
\includegraphics[width=10cm, height=10cm]{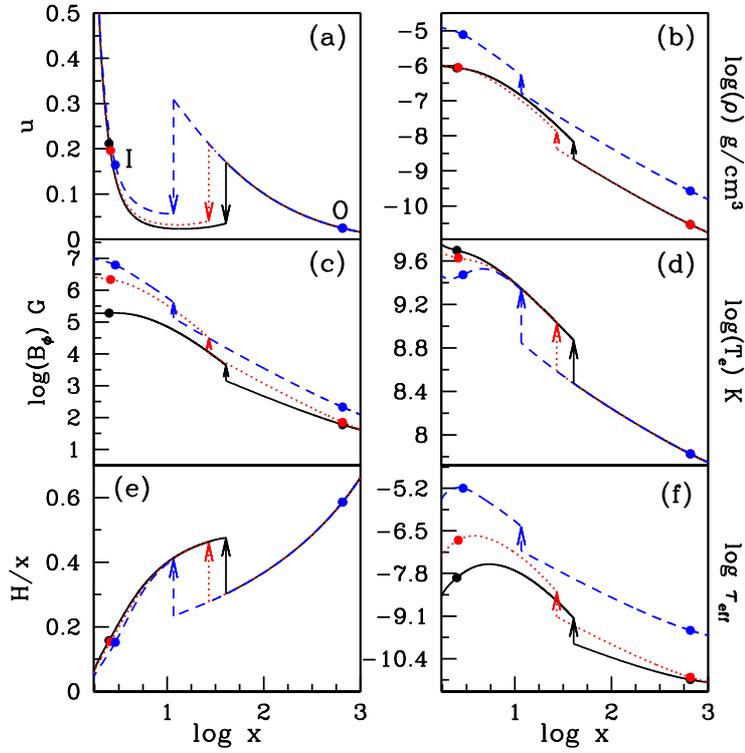}

\caption{Variation of (a) radial velocity, (b) density in g/cm$^3$, (c) the toroidal component of magnetic field in Gauss, (d) the electron temperature in Kelvin, (e) disc aspect ratio ($H/x$) and (f) effective optical depth as a function of logarithmic radial coordinate. The solid, dotted and dashed curves relate to the shocked accretion solutions portrayed in Fig. 1. The critical points are indicated using filled circles where the closer one is the inner critical point and the farthest represents the outer critical point. The position of shock is displayed using vertical arrows. See text for details.}\label{fig:2}
\end{figure}

%%%%%%%%%%%%%%%%%%%%%%%%%%%%%%%%%%%%%%%%%%%%%%%

In Fig. 2, we show the radial disc structure corresponding to shock solutions considered in Fig. 1. 
For this, we consider $M_{\rm BH}$ = 10 $M_{\odot}$ as a standard reference value. Here in 
each panel we show the radial variation of a flow variable where 
solid, dotted and dashed curves depict results corresponding to ($\zeta$$ = 0.6$, $\dot{m} = 0.05$), 
($\zeta$$ = 1.0$, $\dot{m} = 0.05$) and ($\zeta$$ = 1.0$, $\dot{m} = 0.45$), respectively. Moreover, the filled 
dots represent the position of the critical points. In Fig. 2(a) we show the variation of the radial velocity of the flow as a function of distance from the BH. Since the accreting matter experiences greater gravitational pull as it moves closer to the BH, the radial velocity profile steadily increases towards the horizon. Thus, the subsonic flow starting from the outer edge of the disc becomes supersonic as it crosses the outer critical point. This supersonic flow experiences centrifugal barrier in the vicinity of the BH and undergoes a shock jump to become subsonic as indicated by the downward vertical arrow in each of the three cases. The subsonic flow again becomes supersonic after passing through the inner critical point and enters the BH horizon. In Fig. 2(b) we show the variation of density of the accretion flow as a function of radial distance. The density of the flow rises as the flow approaches the horizon and the density is boosted up across the shock transition. The post-shock density rises due to the reduction of flow velocity after the shock jump in order to conserve the mass flux ($\dot{M}$) across the shock transition. This is evident from the sudden rise in density profile in all the cases as indicated by the upward vertical arrow. Next, in Fig. 2(c) we present the variation of the magnetic field strength $B_{\phi}$ as a function of radial distance. For all the cases, we find that $B_{\phi}$ steadily increases with the decrease of radial coordinate. This happens due to the increase of magnetic flux advection rate ($\dot{\Phi}$) with deceasing radius. Increase of $\zeta$ augments $\dot{\Phi}$ and leads to increase of $B_{\phi}$ profile as shown by the dotted curve. Also when $\dot{m}$ rises, there would be an increase of magnetic field lines carried with the matter. So, $B_{\phi}$  profile would also increase as indicated using the dashed curve. Further, $B_{\phi}$ also jumps across the shock transition due to the compression of the flow at the shock location and the conservation of $\dot{\Phi}$ across the shock. In Fig. 2(d), we show the variation of electron temperature ($T_e$) in the flow as a function of radial distance. This is essential since $T_e$ is related to the nature of X-ray spectrum emitted from the disc. We observe a rise in $T_e$ in the flow with decreasing radius for all the cases. $T_e$ is catastrophically augmented across the shock jump showing thereby that the PSC is significantly hotter than the pre-shock region. When $\zeta$ value is increased from $0.6$ to $1.0$ (keeping $\dot{m}$ fixed), there is increase of magnetic field strength that increases the cooling of the electrons due to synchrotron emission. Thus we observe a drop in $T_e$ profile in the inner region (dotted curve). A further drop in the $T_e$ profile is also observed when $\dot{m}$ is 
increased to $0.45$ retaining $\zeta$ fixed (dashed curve). Augmented accretion rate additionally enhances the efficiency of synchrotron emission by the electrons and also causes the drop in $T_e$ profile. In Fig. 2(e), we show the radial variation of the aspect ratio $(H/x)$ of the magnetized accretion disc. We find that the aspect ratio remains below unity throughout the disc for all the cases. Further, due to the high temperature and density in the PSC of the disc, the PSC becomes puffed up and a sudden rise in the disc aspect ratio is observed after the shock jump in all cases. Lastly, in Fig. 2(f), we show the radial distribution of effective optical depth in the vertical direction ($\tau_{\rm eff}$). In the disc, we calculate $\tau_{\rm eff}$ using the relation $\tau_{\rm eff} = \sqrt{\tau_{\rm syn}\tau_{\rm es}}$ 
\citep{Rajesh2010} where, $\tau_{\rm syn}$ denotes the absorption effect arising due to thermal processes and is given by 
$\tau_{\rm syn} =\left( H {\bar{q}}_{\rm syn}/4 \bar{a} T_{e}^4\right)\left(GM_{\rm BH}/c^2\right)$ \citep{Rajesh2010}
where, ${\bar{q}}_{\rm syn}$ is the synchrotron emissivity \citep{Shapiro-Teukolsky83} and $\bar{a}= 
5.6705\times 10^{-5}~{\rm g}~{\rm s}^{-3}~{\rm K}^{-4}$ is the Stefan-Boltzmann 
constant. Here, $\tau_{\rm es}$ denotes the scattering optical depth estimated as $\tau_{\rm es} = \kappa_{\rm es} \rho H$ and where $\kappa_{\rm es} = 0.38~{\rm cm}^2 {\rm g}^{-1}$ is the electron scattering opacity.
Here,  we observe that for all the cases, $\tau_{\rm eff}$ remains well below unity. However, a rise in $\tau_{\rm eff}$ is observed across the shock transition in every case due to greater density at the PSC as observed in Fig. 2(b). When $\zeta$ is increased retaining other parameters fixed, the magnetic activity in the disc is augmented, leading to the increase in $\tau_{\rm syn}$ and the consequent increase of $\tau_{\rm eff}$ profile (dotted curve). Also, when $\dot{m}$ rises to $0.45$ keeping rest of the parameters fixed, there is further rise in $\tau_{\rm eff}$ profile due to increase of density in the accretion flow (dashed curve).   

%%%%%%%%%%%%%%%%%%%%%%%%%%%%%%%%%%%%%%%%%%%%%%%
%%%%%%%%%%%%%%%%% FIG 3 %%%%%%%%%%%%%%%%%
%%%%%%%%%%%%%%%%%%%%%%%%%%%%%%%%%%%%%%
\begin{figure}[ht]
\begin{center}
\includegraphics[width=10cm, height=10cm]{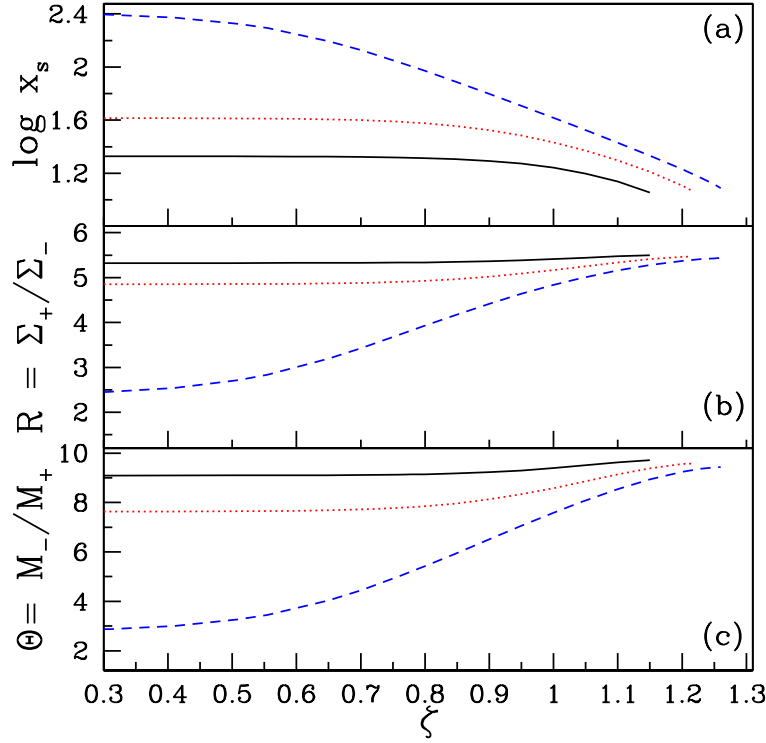}
\end{center}
\caption{Variation of (a) shock location $x_s$, (b) compression ratio $R$, and (c) shock strength $\Theta$ as a function of $\zeta$ for flows injected from $x_{\rm inj} = 1000$ with $\beta_{\rm inj} = 10^5$, $\alpha_T = 0.02$, ${{E}}_{\rm inj}$$=8.9\times10^{-4}$ and $\dot{m} = 0.05$. The Kerr parameter is chosen as $a_s = 0.8$. Solid, dotted and dashed curves represent the results corresponding to $\lambda_{\rm inj}$ = 0.10888$\lambda_K$, $0.11014$$\lambda_K$ and $0.11141$$\lambda_K$ respectively. See text for details.
}
\end{figure}
%%%%%%%%%%%%%%%%%%%%%%%%%%%%%%%%%%%%%%%%%%%%%%%%

One of the pertinent aspect in understanding the magnetically supported accretion flow around the rotating BHs is to study the dependence of the shock properties on the magnetic flux advection rate ($\dot{\Phi}$). For this purpose in Fig. 3, we fix the outer edge of the disc at $x_{\rm inj} = 1000$ and inject matter to accrete with ${{E}}_{\rm inj}$ $= 8.9\times10^{-4}$, $\beta_{\rm inj} = 10^5$, $\alpha_T = 0.02$ and $\dot{m} = 0.05$, respectively. Kerr parameter is held fixed at $a_s = 0.8$. It has already been specified that the $\zeta$$-$parameter decides the advection rate of toroidal magnetic flux ($\dot{\Phi}$). Thus in Fig. 3(a), we present the variation of $x_s$ with $\zeta$ for three sets of specific angular momentum, $\lambda_{\rm inj}$ = 0.10888$\lambda_K$ (solid curve), $\lambda_{\rm inj}$ = 0.11014$\lambda_K$ (dotted curve) and $\lambda_{\rm inj}$ = 0.11141$\lambda_K$ (dashed curve). We observe that for a given $\lambda_{\rm inj}$, shocks form for an ample range of $\zeta$ and when $\zeta$ is increased, shock front moves towards the BH horizon. The increase of $\zeta$ leads to the increase of toroidal magnetic flux advection rate in the radial direction as the flow proceeds towards the BH \citep{oda07,oda12}. With the increase of magnetic activity in the disc, the flow gets efficiently cooled. Since the cooling efficiency is greater in the PSC due to higher temperature and density there, the post-shock pressure drops and the shock front is pushed towards the BH in order to maintain pressure balance across it. The increase of magnetic activity in the disc also enhances the angular momentum transport towards the outer region of the disc. This weakens the centrifugal repulsion against gravity and also contributes to shift the shock towards the horizon. When $\zeta$$-$parameter is increased beyond a critical value ($\zeta$$^{\rm cri}$), the shock conditions are no longer satisfied and standing shock fails to form. Moreover, for a given $\zeta$, we find that when flow is characterized with relatively large $\lambda_{\rm inj}$, shocks form over a large range of distances and vice versa. This clearly indicates that shock transition in BH accretion flows are centrifugally driven.

Now, corresponding to the shock solutions shown in Fig. 3(a), we present the variation of relating shock compression ratio ($R$) and the shock strength ($\Theta$) with $\zeta$$-$parameter in Fig. 3(b) and Fig. 3(c), respectively. Here, $R$ is defined as the ratio of the vertically integrated post-shock density to pre-shock density as $R = \Sigma_{+}/\Sigma_{-}$. On the other hand, $\Theta$ is defined as the ratio of pre-shock  to post-shock Mach number ($\Theta = M_{-}/M_{+}$). Since stronger shocks are located closer to the BH horizon, we find that $R$ steadily increases with $\zeta$ in Fig. 3(b). We find that for shocks closest to the BH horizon, $R\sim 5$. Moreover, similar variational characteristics of $\Theta$ is observed in Fig. 3(c) as in the case of $R$. It is to be noticed that $R$ and $\Theta$ are almost independent with the flux advection rate for the results shown using solid and dotted curves. The reason for this behaviour is the weak angular momentum transport rate for low value of $\zeta$. This claim is consistent with equation (5), where, $\mathbb{T}_{x\phi}$ is proportional to the sum of vertically integrated pressure of the flow ($W$) and the ram pressure ($\Sigma u^2$). Here, $W$ has contribution from the gas pressure ($p_{\rm gas}$) as well as magnetic pressure ($p_{\rm mag}$) in the flow. Now, from Fig. 2(c), we find that when $\zeta$ value is increased, the value of $B_{\phi}$ increases at a particular radial coordinate. Thus, $p_{\rm mag}(=<B_{\phi}>^2/8\pi)$ rises with the increase of $\zeta$ for a given radial coordinate and consequently, $\mathbb{T}_{x\phi}$ also augments at that location. So, $\zeta$ controls the angular momentum transport rate in the flow. The major motivation in this figure is to establish that the accretion flow admits shock for a wide range of angular momentum as well as $\zeta$. The chosen values of $\lambda_{\rm inj}$ for the solid and dotted curves are lower than that of the dashed curve. So for the solid and dotted curves, the shocks form closer to the black hole as compared to the dashed curve. Also when $\zeta$ is low ($\lesssim$ 0.8), the rate of transport of angular momentum transport in the disc is expected to be feeble. Due to the proximity of the shock location to the BH horizon and meagre angular momentum transport rate, there is a small shift in shock location when $\zeta$ is lowered and we do not observe a significant variation in the compression ratio and shock strength for the results shown using solid and dotted curves.

%%%%%%%%%%%%%%%%%%%%%%%%%%%%%%%%%%%%%%%
%%%%%%%%%%%%%%%%% FIG 4 %%%%%%%%%%%%%%%%%%%%%
%%%%%%%%%%%%%%%%%%%%%%%%%%%%%%%%%
\begin{figure}[ht]
\begin{center}
\includegraphics[width=10cm, height=10cm]{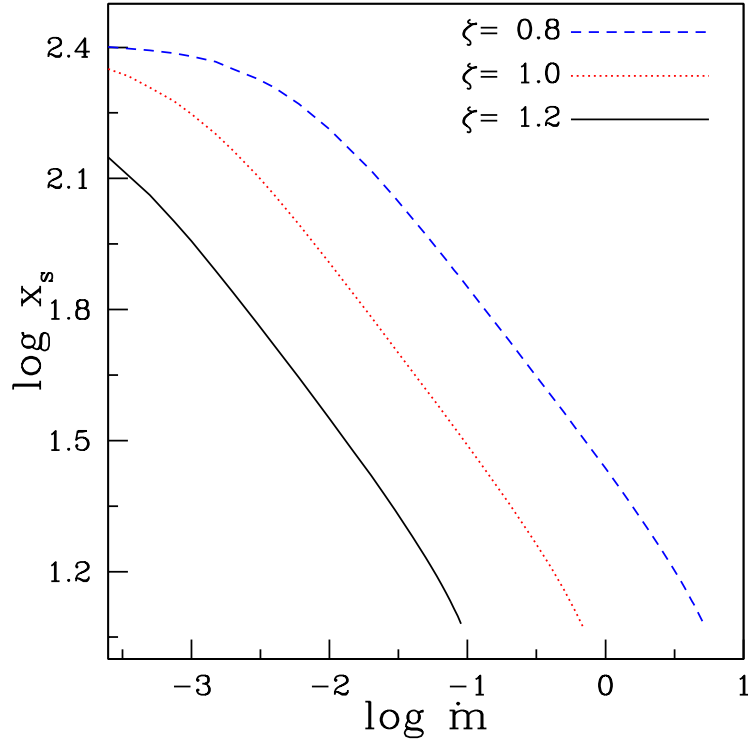}
\end{center}
\caption{Shock location ($x_s$) variation as a function of $\dot{m}$. Flow is injected from the outer edge $x_{\rm inj} = 1000$ with $\beta_{\rm inj} = 10^5$, $\alpha_T = 0.02$, ${{E}}_{\rm inj}$$=8.9\times10^{-4}$ and $\lambda_{\rm inj}$ = 0.11141$\lambda_K$. The spin of the BH is chosen as $a_s = 0.8$. Results plotted with solid, dotted and dashed curves are for $\zeta$$ = 1.2$, $1.0$, and $0.8$, respectively. See the text for details.
}
\end{figure}
%%%%%%%%%%%%%%%%%%%%%%%%%%%%%%%%%%%%%%%%%%%%%%%

In Fig. 4, we show the variation of shock location ($x_s$) as function of $\dot{m}$. At the injection radius $x_{\rm inj} = 1000$, the flow parameters are chosen as ${{E}}_{\rm inj}$$=8.9\times10^{-4}$, $\lambda_{\rm inj}$ = 0.11141$\lambda_K$, $\beta_{\rm inj} = 10^5$ and $\alpha_T = 0.02$, respectively. The BH spin parameter is fixed as $a_s = 0.8$. Here, solid, dotted and dashed curves refer to the results corresponding to $\zeta$$ = 1.2, 1.0$ and $0.8$, respectively. We notice that for a given $\zeta$, $x_s$ decreases with the increase of $\dot{m}$. It is already pointed out that shock front advances towards the BH when $\dot{m}$ is increased (see Fig. 1) and the overall pressure in the PSC is expected to be reduced. Therefore, for a given $\zeta$, we observe an anti-correlation between $x_s$ and $\dot{m}$. When $\dot{m}$ rises beyond a critical limit ($\dot{m}^{\rm cri}$), standing shocks do not from since the shock conditions are not satisfied anymore. Again, $x_s$ is also found to decrease with the increase in $\zeta$, when $\dot{m}$ is fixed. In reality, when $\zeta$ parameter is augmented, synchrotron cooling efficiency in the flow is higher as there is increased magnetic activity in the disc. Due to the efficient cooling at the PSC, the post-shock pressure decreases and the shock location approaches towards the BH in order to maintain the pressure balance across the shock front. Thus, for a given $\dot{m}$, $x_s$ anti-correlates with $\zeta$.

\section{Astrophysical Relevance}
%%%%%%%%%%%%%%%%%%%%%%%%%%%%%%%%%%%%%%%%%%%%%%%%%%%%%%%%%%%%%%%%%%%%%%%%%%%%%%%%%%%
\begin{table*}
%\begin{center}
%\begin{table*}
%\begin{minipage}{100mm}
%\begin{sidewaystable} % <-- HERE
%\begin{minipage}{210mm}
%\begin{sidewaystable}[counterclockwise] % <-- HERE
%\begin{sidewaystable}[clockwise]
\tiny{
\caption {Calculation of luminosity of shock. In Column 1, we make a list of the names of the sources and 
in Column 2-4 we give the corresponding mass, accretion rate and spin. In Column 5-6, we report the model parameters 
and in Column 7-9, we indicate the maximum energy dissipated, the shock location and the estimated 
maximum luminosity of shock obtained using equation (15). The core radio luminosity values reported 
from observations are listed in column 10.}
%\begin{center}
\renewcommand{\arraystretch}{1.0}
\begin{tabular}{l c c c c c c c c ccccc}
\hline\hline\\
Object & $M_{\rm BH}$ & $\dot{m}$ & $a_s$ & ${E}_{\rm in}$ & $\lambda_{\rm in}$ & $\Delta {E}^{\rm max}$ & $x_s$ &
$\mathcal{L}^{\rm max}_{\rm shock}$ & $\mathcal{L}^{\rm Obs}_{\rm jet}$\\

&$({\rm M}_{\odot})$ & $({\dot M}_{\rm Edd})$ & & $(10^{-2})$ & $(\lambda_K)$ & ($10^{-2} c^2$) & $(r_g)$ & (erg s$^{-1}$) & (erg s$^{-1}$) \\ \hline\\
 A0620-00  & $6.60$$~^a$ &  $5.0 \times 10^{-4}$$~^b$& $0.12$$~^c$& $-0.6517$  & $0.88972$& $0.8754$ &  $16.54$ &  $3.61 \times 10^{33}$ & $1.0 \times 10^{33}$$~^{\bigstar}$ \\\\
 LMC X-3  & $6.98$$~^d$ &  $2.365$$~^e$& $0.25$$~^f$& $-1.5036$  & $0.85117$ & $1.5642$ &  $16.03$ &  $3.23 \times 10^{37}$ & $-$ \\\\
 H1743-322  & $11.21$$~^{g}$ &  $0.127$$~^{h}$& $0.4$$^{i}$& $-2.2296$  & $0.76128$ & $2.2488$& $15.22$ &  $4.0 \times 10^{36}$ & $3.60 \times 10^{38}$$~^{\bigstar}$ \\\\
 XTE J1550-564  & $9.10$$~^j$ &  $0.528$$~^k$& $0.49$$~^k$& $-2.4323$  & $0.68529$ & $2.4216$ &  $14.27$ &  $1.45 \times 10^{37}$ & $3.41 \times 10^{38}$$~^l$ \\\\
 GRO J1655-40  & $5.31$$~^m$ &  $0.596$$~^n$& $0.7^{o}$& $-2.8775$  & $0.40495$ & $2.7427$& $13.25$ &  $1.08 \times 10^{37}$ & $1.95 \times 10^{36}$$~^{p}$ \\\\
 XTE J1118+480$^{\dag}$  & $7.50$$~^q$ &  $3.0 \times 10^{-4}$$~^b$& $0.9$$^{r}$& $-2.9702$  & $0.29669$ & $3.0866$ &  $9.07$ &  $8.68 \times 10^{33}$ & $4.0 \times 10^{32}$$~^{\bigstar}$ \\\\
 LMC X-1$^{\dag}$  & $10.91$$~^s$ &  $0.884$$~^t$& $0.92$$~^t$& $-4.1158$  & $0.26757$ & $3.9125$ &  $8.972$ &  $4.71 \times 10^{37}$ & $2.0 \times 10^{39}$$~^u$ \\\\
 Cyg X-1$^{\dag}$  & $14.8$$~^v$ &  $0.061$$~^w$& $0.97$$~^{x}$& $-3.3274$  & $0.35534$ & $3.4323$& $8.02$ &  $3.87 \times 10^{36}$ & $1.0 \times 10^{37}$$~^{y}$ \\\\
 GRS1915+105$^{\dag}$  & $10.1$$~^z$ &  $3.188$$~^{zx}$& $0.98$$~^{zy}$& $-3.5997$  & $0.36401$ & $3.4312$& $8.02$ &  $1.38 \times 10^{38}$ & $1.0 \times 10^{38}$$~^{zx}$ \\\\
\hline
\end{tabular}
%\end{center}
%\footnotesize{
\begin{quote}{\bf References.}
$^a$\citet{Cantrell2010}, $^b$\citet{Yang2016}, $^c$\citet{Gou2010}, 
$^d$\citet{Orosz2014}, $^e$\citet{Kubota-etal-2010}, $^f$\citet{Steiner2014}, 
$^{g}$\citet{Molla-etal-2017}, $^{h}$\citet{Bhattacharjee-etal-2017}, $^{i}$\citet{Tursunov2018}, 
$^j$\citet{Orosz2011a}, $^k$\citet{Steiner2011}, $^l$\citet{Fender2004}, 
$^m$\citet{Motta-etal-2014}, $^n$\citet{Luketic-etal-2010}, $^{o}$\citet{Shafee_2006},
$^p$\citet{Migliari-etal-2007}, $^q$\citet{Khargharia2013}, $^r$\citet{Chaty_2003}, 
$^s$\citet{Orosz2009}, $^t$\citet{Gou2009}, $^u$\citet{Cooke2007}, $^v$\citet{Orosz2011b}, 
$^{w}$\citet{Gou2011}, $^{x}$\citet{Fabian2012}, $^{y}$\citet{Gallo2005}, $^{z}$\citet{Steeghs2013},
$^{zx}$\citet{Belloni2000}, $^{zy}$\citet{Miller-etal-2013}.\\
%$^{**}$ Due to unavailability of the estimation of spin value for XTE J1118+480 in the literature, we assume the source to be a non-rotating BH.\\
$^{\bigstar}$ $L^{\rm Obs}_{\rm jet} = \dot{M}_{\rm out}c^2$ is used for the sources A0620-00, XTE J1118+480 and H 1743-322, where $\dot{M}_{\rm out}$ is the outflow rate. For the sources A0620-00 and XTE J1118+480, the outflow rate is adopted from \citet{Yang2016}. For H1743-322, $\dot{M}_{\rm out}$ is considered following \citet{Miller-etal-2006}.

\end{quote}
}
\bigskip
\bigskip
%\end{minipage}
%\end{sidewaystable} % <-- HERE
%\end{minipage}
%\end{table*}
%\end{center}
\end{table*}
%%%%%%%%%%%%%%%%%%%%%%%%%%%%%%%%%%%%%%%%%%%%%%%%%%%%%%%%%%%%%%%%%%%%%%%%%%%%%%%%%%%%%%%%%%%

Thus far, we have dealt with non dissipative shocks where the specific energy essentially 
remains conserved across the shock front \citep{Chak89}. However, as seen in Fig. 2(f), 
the sudden compression of the flow at the shock causes the optical depth of the flow to 
rise significantly across the shock front. Since the shock is thin and the optical 
depth there is very high, thermal Comptonization of the photons is expected to be very 
important. \citet{ct95} pointed out that due to the thermal Comptonization process, the 
dissipation of energy of the accretion flow at the shock is likely. This energy 
dissipation mechanism eventually reduces the temperature of the flow in the PSC and 
the energy loss is proportional to the difference in temperature in the post- and pre-shock 
flows. Following this, the loss of energy ($\Delta {{E}}$) at the shock is assessed 
as \citep{Das-etal10}, 
$$ 
\Delta E = f_dn (c_{s+}^{2} - c_{s-}^{2}), 
\eqno(14) 
$$ 
where $c_{s-}$ and $c_{s+}$ denote the pre-shock and post-shock sound speeds, 
respectively. Here, $f_d$ refers to the fraction of thermal energy difference 
dissipated across the shock front, which is treated as a free parameter in this work \citep{Das-etal10,
Singh_Chakrabarti_2011,sarkar_2013,Kumar2013,Sarkar-et-al2018,Das-Sarkar2018}. The kinetic power lost from the disc can be evaluated by regarding the shock luminosity as in \citet{Le_Becker2004,Le_Becker2005,Sarkar-Das2016,Sarkar-et-al2018}, 
$$ 
\mathcal{L}_{\rm total}=\mathcal{L}_{\rm shock}=\dot M \times \Delta E \times c^2~~{\rm erg~s^{-1}}, 
\eqno(15) 
$$ 
where, $\mathcal{L}_{\rm total}$, $\mathcal{L}_{\rm shock}$ and $\dot M$ denote the kinetic power lost by the disc, the 
luminosity of shock and the accretion rate for a specified source, respectively. The energy loss at the shock front due 
to Comptonization ($\Delta {{E}}$) is essentially a function of the number of soft photons and the number density 
of electrons \citep{Mondal-etal-2014}. In this paper, we fix $f_d = 0.998$ all throughout, 
for the sake of representation.

Subsequently, we compute the maximum luminosity of shock ($\mathcal{L}^{\rm max}_{\rm shock}$) 
relating to maximum dissipation of energy at the shock. Here, $\alpha_B=0.001$ and $\beta_{\rm in} = 10^5$ are considered for all cases. Also, in \citet{oda07}, the authors have obtained radial structure of the accretion disc considering $\zeta= 1$, which remains optically thin throughout. For $\zeta \sim 0$, the authors showed that steady global transonic solutions with magnetic fields were not obtained beyond a certain critical accretion rate. They further reported that for $\zeta \sim 1$, accretion solutions connecting the outer boundary to the inner region can be obtained for large accretion rates with low$-\beta$ region at the inner part of the disc. Moreover, \citet{oda12} delineated that for small values of $\zeta$ (e.g., $\zeta \sim 0$), it is not possible to obtain the magnetically supported low$-\beta$ disc solutions but the usual advection dominated accretion flow (ADAF)/ radiatively inefficient accretion flow (RIAF) solutions can be explained. Further in Table 1 of \citet{oda12}, the authors showed the formation of low$-\beta$ disc in global accretion solutions for $0.5\leq \zeta \leq 1$, by suitably choosing the other disc parameters. For, $\zeta > 1$, the authors additionally showed that magnetic pressure would dominate in the accretion flow even at low accretion rates. Considering all these, we argue that $\zeta \geq 0.5$ is a reasonable approximation in the accretion disc to obtain the low$-\beta$ solutions and this can support the observational values. Thus, we choose $\zeta = 1$ as a representative value in the analysis that follows. 

In Table 1, we present the physical parameters 
of the stellar mass BH sources including model parameters and compute the maximum luminosity of shock. Columns 
1-4 of Table 1 show the names of sources, their mass ($M_{\rm BH}$), accretion rate (${\dot m}$) 
and spin ($a_s$). In section 5-6, we note the specific energy at the inner critical point (${{E}}_{\rm in}$) and the angular momentum at the inner critical point ($\lambda_{\rm in}$). In column 7, we note the maximum dissipated energy $\Delta {E}^{\rm max}$ and in column 8, we indicate the position of shock 
($x_s$). In column 10, the maximum luminosity of shock $\mathcal{L}^{\rm max}_{\rm shock}$ 
is reported. Our primary objective in this examination is to estimate the upper limit of energy that 
can be removed from the PSC to power the outflowing matter from the disc as 
Jets. Accordingly, we evaluate the maximum dissipated energy $\Delta {E}^{\rm max}$ at 
the position of shock jump. For the stellar mass BH sources under consideration, we find that 
the evaluated luminosities of shock are in consensus with the core radio luminosity values
 $\mathcal{L}^{\rm Obs}_{\rm jet}$ (in column 10) reported from observations 
\citep{Fender2004,Cooke2007,Gallo2007,Migliari-etal-2007,Belloni2000}, subject to availability. To estimate the observed jet luminosity ($L_{\rm jet}^{\rm Obs}$) from the outflow rate ($\dot{M}_{\rm out}$), we have used the relation $L_{\rm jet}^{\rm Obs} = \dot{M}_{\rm out}c^2$. Here, we assume 100\% energy conversion efficiency on the basis of the common definition $L_{\rm Edd} = \dot{M}_{\rm Edd} c^2$ for the Eddington limit on the mass accretion rate used by many authors \citep{Rezzolla-2013,Straub-et-al-2014,Yuan2014}. Although 100\% energy conversion efficiency is unphysical, however, the definition used would give us the maximum observed luminosity achievable from the outflow rate available in the literature, which is compared with the calculated maximum shock luminosity from our model. For the BH sources marked with superscript $\dag$, the reported values of spin from the literature exceeds $0.8$. In such a case, it is inappropriate to use the \citet{Chak-Mondal2006} potential to calculate the shock luminosity corresponding to these sources. Nevertheless, we carry out the analysis using this potential in order to get a qualitative estimate of $\mathcal{L}^{\rm max}_{\rm shock}$ for these sources.

\section{Summary}

We study the effects of the variation of magnetic flux advection rate ($\dot{\Phi}$) and accretion rate ($\dot{m}$) on the properties of shock in accretion flow around a rapidly rotating BH ($a_s = 0.8$). In the global transonic accretion solutions, heating takes place due to the thermalization of magnetic energy, while cooling occurs via synchrotron emission. In Fig. 1, we present the combined effect of 
variation of $\dot{m}$ and $\zeta$$-$parameter on the shock dynamics. For fixed injection parameters at the outer edge, we find that the shock location shifts towards the horizon with the increase of $\zeta$ (via the corresponding increase of $\dot{\Phi}$) as well as the increase of $\dot{m}$. In Fig. 2, we have demonstrated the radial variation of various flow variables corresponding to the global shocked accretion solutions considered in Fig. 1. We observe that, the PSC is hotter, denser, optically thick and also contains higher magnetic field strength as compared to the pre-shock flow. In Fig. 3, we study the effect of the increase of $\zeta$ on the shock dynamics. 
The increase of $\zeta$ leads to the increase of toroidal magnetic flux advection rate in the radial direction as 
the flow gets closer to the BH. With the increase of magnetic activity in the disc, the flow is efficiently 
cooled. Since the cooling efficiency is greater in the PSC due to higher temperature and density 
there, the post-shock pressure drops and the shock front is pushed towards the BH in order to maintain pressure 
balance across it. The increase of magnetic activity in the disc (due to increase of $\zeta$) also enhances the 
angular momentum transport towards the outer region of the disc. This further weakens the centrifugal repulsion against 
gravity and also contributes to shift the shock towards the horizon. As a consequence, the shock compression 
ratio ($R$) and the shock strength ($\Theta$) also rise (Fig. 3 (b) \& 3(c)). Furthermore, from Fig. 3 we also 
observe that global accretion solutions with shock exist for an ample range of $\zeta$$-$parameter and angular momentum. Next in Fig. 4, 
we explore the effect of increase of $\dot{m}$ on the shock position. For a given set of injection parameters 
at the outer edge, we observe that global shocked accretion solutions 
exist for a wide range of $\dot{m}$. However, the range of $\dot{m}$ for shock anti-correlates with the increase 
of $\zeta$. Finally in Section 4, we consider the astrophysical application of the present formalism. We 
estimate the maximum energy dissipated across the shock front via thermal Comptonization phenomenon 
and the corresponding shock luminosity ($\mathcal{L}^{\rm max}_{\rm shock}$) for several 
stellar mass BHCs. We find that the estimated $\mathcal{L}^{\rm max}_{\rm shock}$ is in well agreement 
with the observed values of jet kinetic luminosity ($\mathcal{L}^{\rm Obs}_{\rm jet}$). If the value of 
$\mathcal{L}^{\rm Obs}_{\rm jet}$ is unavailable for a given source, the present model can predict the 
maximum shock luminosity $\mathcal{L}^{\rm max}_{\rm shock}$ that may be observed.  

In the end, we would like to mention the limitations of the present formalism. For the sake of simplicity, 
we have adopted the \citet{Chak-Mondal2006} pseudo-Kerr potential to mimic the general relativistic effects 
around a rotating BH. The potential gives satisfactory results till spin value $a_s\leq 0.8$. Thus the accretion 
flow properties around an extremely rotating BH cannot be explored satisfactorily in the present approach. Although, 
we have attempted to calculate $\mathcal{L}_{\rm max}^{\rm shock}$ for four stellar mass BH candidates with 
$a_s > 0.8$, this provides only a qualitative estimate. Again, the 
magnetic fields are likely to cause the generation of jets and outflows from the disc, which have not been 
considered in the present work. Moreover, the adiabatic index is considered to be a global constant in the 
paper, although it should be calculated self consistently from the global accretion solution. Since the magnetized accretion flow in this work remains optically thin throughout (Fig. 2f), Comptonization of the soft synchrotron photons that are produced locally would be important \citep{Yuan2014}. However, to carry out a simplistic treatment, we have excluded the cooling effect due to Comptonization in this work. \citet{Ipser1983} have indicated that the radiative cooling efficiency in the accretion flow is enhanced due to the Comptonization process. Again, recently \citet{Dihingia-etal2018} have studied two-temperature shocked accretion flows around BHs including the effect of Comptonization. In this work, the authors showed that due to the rise in density of the PSC as a result of shock transition, all the radiative processes in the flow viz; bremsstrahlung, synchrotorn and Comptonization due to synchrotorn photons are very much efficient in the PSC. Thus for fixed outer boundary conditions, if Comptonization due to synchrotorn photons is included in our present work, we expect that the radiative cooling in the PSC would be enhanced and the shock location would shift towards the horizon in order to maintain the pressure balance across the shock. This will result in a quantitative change in the related shock properties as reported in the present work. For example, the shock compression ratio ($R$), shock strength ($\Theta$) and the critical limits of $\zeta$ and $\dot{m}$, as found in Figs. (3) and (4), would be quantitatively modified. However, the qualitative behaviour of the results is expected to remain unaltered. We would like to address these issues in future.

\section{Acknowledgements}

BS acknowledges Dr. Ramiz Aktar and Mr. Indu Kalpa Dihingia for useful discussions on specific topics of the 
paper. AR acknowledges a Commonwealth Rutherford Fellowship. Authors would like to thank the referee for his valuable comments and constructive suggestions, which helped to improve the presentation of the paper.
%the anonymous referee for her or his constructive comments and suggestions, which helped us to carry out this research deeply and improve the presentation of this paper. 

% Don't change these lines
%\bsp	% typesetting comment
\label{lastpage}

\end{document}